\documentclass[onecolumn,10pt,showpacs]{revtex4}
\usepackage{graphicx}
\usepackage{epsfig}
\usepackage{bm}
\usepackage{amsfonts}

\input epsf

\begin{document}

\title{\Large Observational Constraints of New Variable Modified\\ Chaplygin Gas Model}

\author{\bf Jhumpa Bhadra$^1$\footnote{bhadra.jhumpa@gmail.com}
and Ujjal Debnath$^2$\footnote{ujjaldebnath@yahoo.com ,
ujjal@iucaa.ernet.in}}

\affiliation{$^1$Department of Mathematics, Heritage Institute of
Technology, Anandapur, Kolkata-700 107, India.\\
$^2$Department of
Mathematics, Bengal Engineering and Science University, Shibpur,
Howrah-711 103, India.\\}

\date{\today}

\begin{abstract}
Assuming the flat FRW universe in Einstein's gravity filled with
New Variable Modified Chaplygin gas (NVMCG) dark energy and dark
matter having negligible pressure.  In this research work we
analyze the viability on the basis of recent observation. Hubble
parameter $H$ is expressed in terms of the observable parameters
$H_0$, $\Omega_m^0$ and the model parameters $A_0$, $B_0$, $C_0$,
$m$, $n$, $\alpha$ and the red shift parameter $z$. Here we find a
best fitted parameter range of $A_0$, $B_0$ keeping $0\leq \alpha
\leq 1$ and using Stern data set (12 points) by minimizing the
$\chi^2$ test at 66\%, 90\% and 99\% confidence levels. Next we do
the joint analysis with BAO and CMB observations. Again evaluating
the distance modulus $\mu(z)$ vs redshift ($z$) curve obtained in
the model NVMCG with dark matter with the best fitted value of the
parameters and comparing with that derived from the union2
compilation data.
\end{abstract}


\maketitle

\section{Introduction}

Modern observations like high redshift survey of SNe Ia
\cite{Perlmutter1,Perlmutter2,Riess,Tonry}, CMBR
\cite{Melchiorri,Lange,Jaffe,Netterfield,Halverson}, WMAP
\cite{Briddle,Bennet,Hinshaw,Kogut,Spergel} illustrate the
accelerating phase of the universe. Accelerating phase phenomenon
of the universe is not understood by the standard Big Bang model
of cosmology giving a time like singularity in the past.
Cosmological constant $\Lambda$ with the equation of state
$p=-\rho$ is the simplest candidate in Einstein's gravity which
describe the present accelerating phase of the universe. The
aspects of inflation and the cosmological constant are not
understood well. There are various candidates to play the role of
the dark energy having negative pressure and describe the
accenting view of the present observation. The mystifying fluid
namely dark energy is understood to dominate the 70\% of the
Universe which violets the strong energy condition and 30\% dark
matter (cold dark matters plus baryons). Various effective
candidate of dark energy namely Chaplygin gas with equation of
state $p=-\frac{B}{\rho}$, $B>0$ introduced by \cite{Kamenshchik}.
Furthermore it has been generalized to the form
$p=-\frac{B}{\rho^\alpha}$, $0\leq \alpha \leq 1$ \cite{Gorini}
and modified to $p=A \rho-\frac{B}{\rho^\alpha}$
\cite{Benaoum,Debnath}. The another candidate of dark energy was
introduced by Chakraborty et al \cite{Chakraborty}, known as {\it
New Variable modified Chaplygin Gas} (NVMCG) which follows the
equation $p=A(a)-\frac{B(a)}{\rho^\alpha}$, $0 \leq \alpha \leq
1$, $a$ being the scale factor which gives interesting physical
significance.
\\

For flat universe having the energy densities for dust like matter
and dark energies we need to know the value of critical energy
density and $H(z)$ at high accuracy \cite{Choudhury}. The MCG best
fits with the 3 year WMAP and the SDSS data with the choice of
parameters $A = 0.085$ and $\alpha = 1.724$ \cite{Lu} which are
improved constraints than the previous ones $-0.35 < A < 0.025$
\cite{Jun}. Here we have assumed the new variable modified
Chaplygin gas model in flat FRW cosmology. The joint data analysis
of stern data set with BAO and CMB have been analyzed for this
model. The distance modulus vs redshift has been examined for our
model via redshift-magnitude observational data from Supernova
type Ia (Union 2).

\subsection{Basic Equations for Einstein Gravity}

The Friedmann-Robertson-Walker(FRW) metric is considered as

\begin{eqnarray}
ds^2=-c^2 dt^2+ a^2(t)\left[\frac{dr^2}{1-kr^2}+r^2\left(d\theta^2+\sin^2 \theta d\phi^2\right)\right]
\end{eqnarray}

where $a(t)$ is the scale factor of the universe and $r$, $\theta$, $\phi$ are the dimensionless comoving co-ordinates,
$k$ be the curvature parameter of space-time metric and takes the values $k=0, +1, -1$ for flat, closed, open universe respectively.\\

We consider a spatially flat universe ($k=0$) with dark energy
and dark matter (non-interacting). Thus the Einstein equations becomes (choosing $8\pi G=c=1$)

\begin{eqnarray} H^2 = \frac{1}{3} \rho \end{eqnarray}
\begin{eqnarray} \dot{H}=-\frac{1}{2}\left(p + \rho \right)\end{eqnarray}

where $\rho=\rho_{DE}+\rho_m$ the total energy density of the
universe and $p=p_{DE}+p_{m}$ the total pressure. $\rho_{DE}$,
$p_{DE}$ are the energy density and pressure for dark energy
respectively and $\rho_{m}$, $p_{m}$ that for dark matter. For
non-interacting fluid conservation equations become

\begin{eqnarray} \dot{\rho}_{DE}+3H (\rho_{DE}+p_{DE})=0 \end{eqnarray}
and
\begin{eqnarray} \dot{\rho}_{m}+3H (\rho_{m}+p_{m})=0 \end{eqnarray}

We have assumed that the universe is filled with New variable
modified Chaplygin Gas (NVMCG) as dark energy whose EoS is
\cite{Chakraborty}
\begin{eqnarray}
p=A(a)\rho_{DE}-\frac{B(a)}{\rho_{DE}^\alpha} ~~~~ \mbox{with } 0\leq \alpha \leq 1
\end{eqnarray}

where $A(a)$, and $B(a)$ are function of the scale factor $a$. In
particular, choose $A(a)=A_0 a^{-n}$ and $B(a)=B_0 a^{-m}$ with
$A_0$, $B_0$, $m$, $n$ are positive constants. For $n=m=0$, this
model reduces to modified Chaplygin Gas and for $n=0$, the model
reduces to the variable modified Chaplygin gas model.\\

Expression for the energy density for NVMCG model is obtained from
(4) as \cite{Chakraborty}

\begin{eqnarray}
\rho=a^{-3}\exp \left({\frac{3A_0 a^{-n}}{n}}\right) \left[C_0
+\frac{B_0}{A_0} \left(\frac{3A_0
(1+\alpha)}{n}\right)^{\frac{3(1+\alpha)+n-m}{n}} \Gamma
\left(\frac{m-3(1+\alpha)}{n},
\frac{3A_0(1+\alpha)}{n}a^{-n}\right)\right]^{\frac{1}{1+\alpha}}
\end{eqnarray}

where $C_0$ is an integration constant and $\Gamma (s,t)$ is the upper incomplete gamma function.\\

For dark matter, the EoS is $p_{m}=0$ and so $\rho_m= \rho_{m}^0
a^{-3}$, $\rho_{m}^0=3 \Omega_{m}^0 H_0$. The Hubble parameter
($H$) in terms of redshift parameter ($z$) can be expressed as
(from eq. (2))
\begin{eqnarray*}
3 H^2=(1+z)^3 \left[3 H_0 \Omega_{m}^0+\exp \left({\frac{3A_0 \left(\frac{1}{1+z}\right)^{-n}}{n}}\right)\left\{C_0
+\frac{B_0}{A_0} \left(\frac{3A_0
(1+\alpha)}{n}\right)^{\frac{3(1+\alpha)+n-m}{n}}\right.\right.
\end{eqnarray*}
\begin{eqnarray}
\left.\left. \times \Gamma
\left(\frac{m-3(1+\alpha)}{n},
\frac{3A_0(1+\alpha)}{n}\left(\frac{1}{1+z}\right)^{-n}\right)\right\}^{\frac{1}{1+\alpha}}\right]
\end{eqnarray}

Subsequently, we investigate the bound on the model parameter by
observational data fitting. The parameters are determined by
$H(z)$-$z$ (Stern), BAO and CMB data analysis
\cite{Wu,Paul1,Paul2,Paul3,Paul4} using $\chi^2$ minimization
technique from Hubble-redshift data set.

\section{Observational data analysis $H(z)$-$z$ (Stern), BAO and CMB Data as a Constraining Tool}

From the above expression we can write, the Hubble parameter
$H(z)$ which can be put in the form as

\begin{eqnarray}
H^2(A_0, B_0, \alpha, m, n, C_0, z)=H_0^2 E^2(A_0, B_0, \alpha, m,
n, C_0, z)
\end{eqnarray}

where,
\begin{eqnarray*}
E(A_0, B_0, \alpha, m, n, C_0, z)=\frac{(1+z)^{3/2}}{H_0\sqrt{3}}
 \left[3 H_0 \Omega_{m}^0+\exp \left({\frac{3A_0 \left(\frac{1}{1+z}\right)^{-n}}{n}}\right)\times\right.
\end{eqnarray*}

\begin{eqnarray}
\left.\left\{C_0
+\frac{B_0}{A_0} \left(\frac{3A_0
(1+\alpha)}{n}\right)^{\frac{3(1+\alpha)+n-m}{n}} \Gamma
\left(\frac{m-3(1+\alpha)}{n},
\frac{3A_0(1+\alpha)}{n}\left(\frac{1}{1+z}\right)^{-n}\right)\right\}^{\frac{1}{1+\alpha}}\right]^{1/2}
\end{eqnarray}

Now $E(A_0, B_0, \alpha, m, n, C_0, z)$ contains six unknown
parameters $A_0,B_0,C_0,m,n$ and $\alpha$. Now we will fixing two
parameters and by observational data set the relation between the
other two parameters will obtain and find the bounds of the
parameters.

\begin{center}
\begin{tabular}{|l|}
\hline\hline ~~$z~~Data$~~~~~$H(z)$~~~~~~~~~~~$\sigma$\\ \hline
~~0.00~~~~~~~~~~~73~~~~~~~~~~~$\pm$ 8.0\\
~~0.10~~~~~~~~~~~69~~~~~~~~~~~$\pm$ 12.0\\
~~0.17~~~~~~~~~~~83~~~~~~~~~~~$\pm$ 8.0\\
~~0.27~~~~~~~~~~~77~~~~~~~~~~~$\pm$ 14.0\\
~~0.40~~~~~~~~~~~95~~~~~~~~~~~$\pm$ 17.4\\
~~0.48~~~~~~~~~~~90~~~~~~~~~~~$\pm$ 60.0\\
~~0.88~~~~~~~~~~~97~~~~~~~~~~~$\pm$ 40.4\\
~~0.90~~~~~~~~~~~117~~~~~~~~~~$\pm$ 23.0\\
~~1.30~~~~~~~~~~~168~~~~~~~~~~$\pm$ 17.4\\
~~1.43~~~~~~~~~~~177~~~~~~~~~~$\pm$ 18.2\\
~~1.53~~~~~~~~~~~140~~~~~~~~~~$\pm$ 14.0\\
~~1.75~~~~~~~~~~~202~~~~~~~~~~$\pm$ 40.4
\\\hline\hline
\end{tabular}
\end{center}

\begin{center}
Table 1: The Hubble parameter $H(z)$ and the standard error
$\sigma(z)$ for different values of redshift z.
\end{center}

\subsection{Analysis of $H(z)$-$z$ (Stern) data set}

For given $\alpha, m, n, C_0, z$, $A_0$ and $B_0$ can be best
fitted by minimizing $\chi^2_{H-z}$ given by

\begin{eqnarray}
\chi^2_{H-z}(A_0, B_0, \alpha, m, n, C_0, z)=\sum
\frac{\left(H(A_0, B_0, \alpha, m, n, C_0,
z)-H_{obs}(z)\right)^2}{\sigma_z^2}
\end{eqnarray}

where $H_{obs}$ is the observed Hubble parameter at redshift $z$
and $\sigma_z$ is the error associated with that particular
observation (see table 1) and $H$ represents the theoretical
values of Hubble parameter calculated for our model. Here we use
the observed value of Hubble parameter at different redshifts
(twelve data points) listed in observed Hubble data by Stern et al
\cite{Stern} to analyze our model. We consider the present value
of Hubble parameter $H_0 = 72 \pm 8 Kms^{-1} Mpc^{-1}$ and a fixed
prior distribution. By fixing the model parameter $\alpha \in
[0,1]$, $m,~n$ and $C_0$, we determine the range of other two
parameters $A_0$ and $B_0$ by minimizing (11). The probability
distribution function in terms of the parameters $A_0$, $B_0$,
$C_0$, $m,~n$ and $\alpha$ is given by

\begin{eqnarray}
L=\int e^{-\frac{1}{2}}\chi^2_{H-z} P(H_0) dH_0
\end{eqnarray}

where $P(H_0)$ is the prior distribution function for $H_0$. Now
our best fit analysis with Stern observational data support the
theoretical range of the parameters. In figures 1 and 2, we plot
the graphs for different confidence levels 66\% (solid, blue),
90\% (dashed, red) and 99\% (dashed, black) contours for
$\alpha=0.0001$ and $0.01$ respectively and by fixing the other
parameters. The best fit values of $A_{0}$, $B_{0}$ and the
minimum values of $\chi^{2}$ are tabulated in Table 2.

\[
\begin{tabular}{|c|c|c|c|}
\hline
  ~~~~~~$\alpha$ ~~~~~& ~~~~~~~$A_{0}$ ~~~~~~~~& ~~~$B_{0}$~~~~~&~~~~~$\chi^{2}_{min}$~~~~~~\\
  \hline

  $0.0001$ & 0.0000869 & 2.988 & 32.164 \\
  $~~0.01$ & 0.0000868 & 3.282 & 31.925 \\
   \hline
\end{tabular}
\]
{\bf Table 2:} $H(z)$-$z$ (Stern): The best fit values of $A_{0}$,
$B_{0}$ and the minimum values of $\chi^{2}$ for $m=7$, $n=13$,
$C_0=0.1$ and for different values of $\alpha$.

\begin{figure}
\epsfxsize = 3 in \epsfysize = 2.5 in \epsfbox{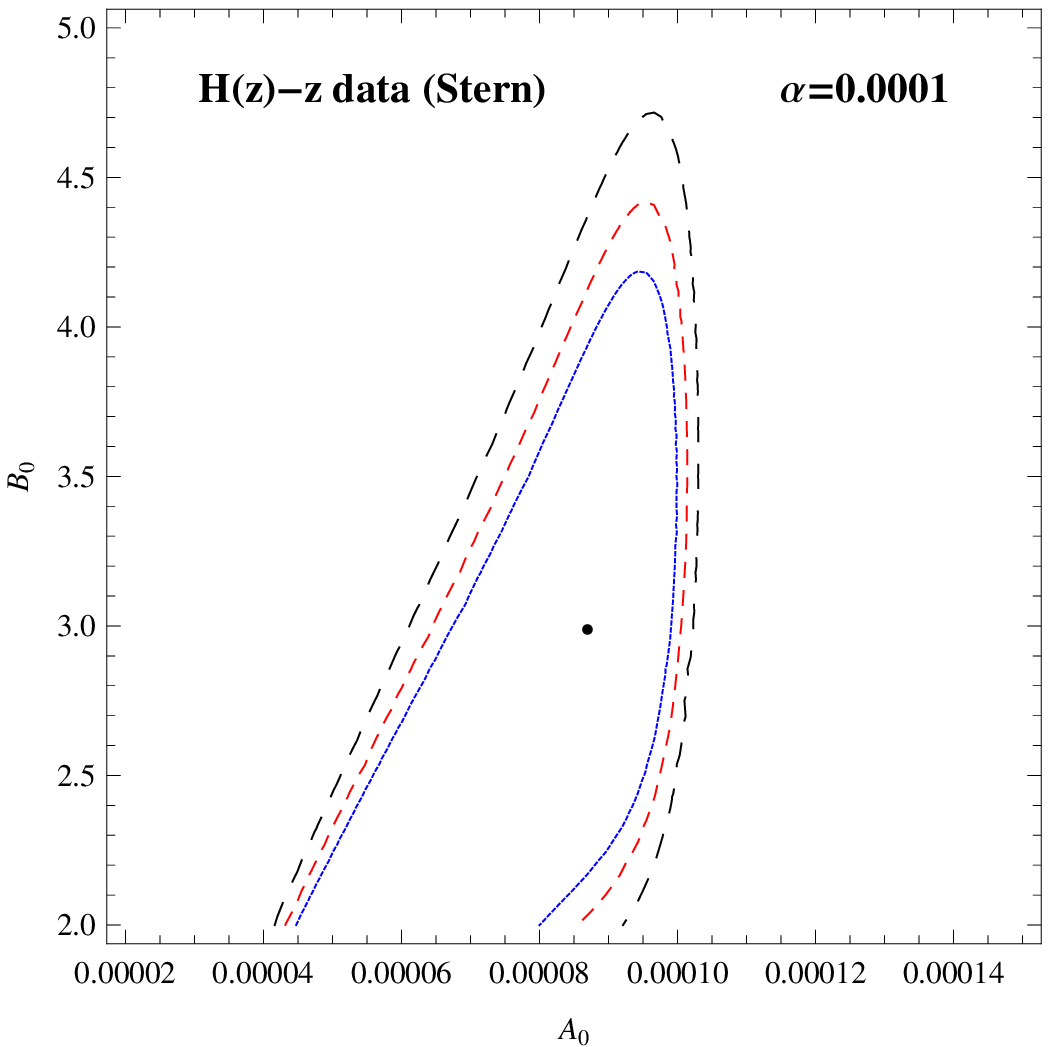}~~~~~
\epsfxsize = 3 in \epsfysize = 2.5 in \epsfbox{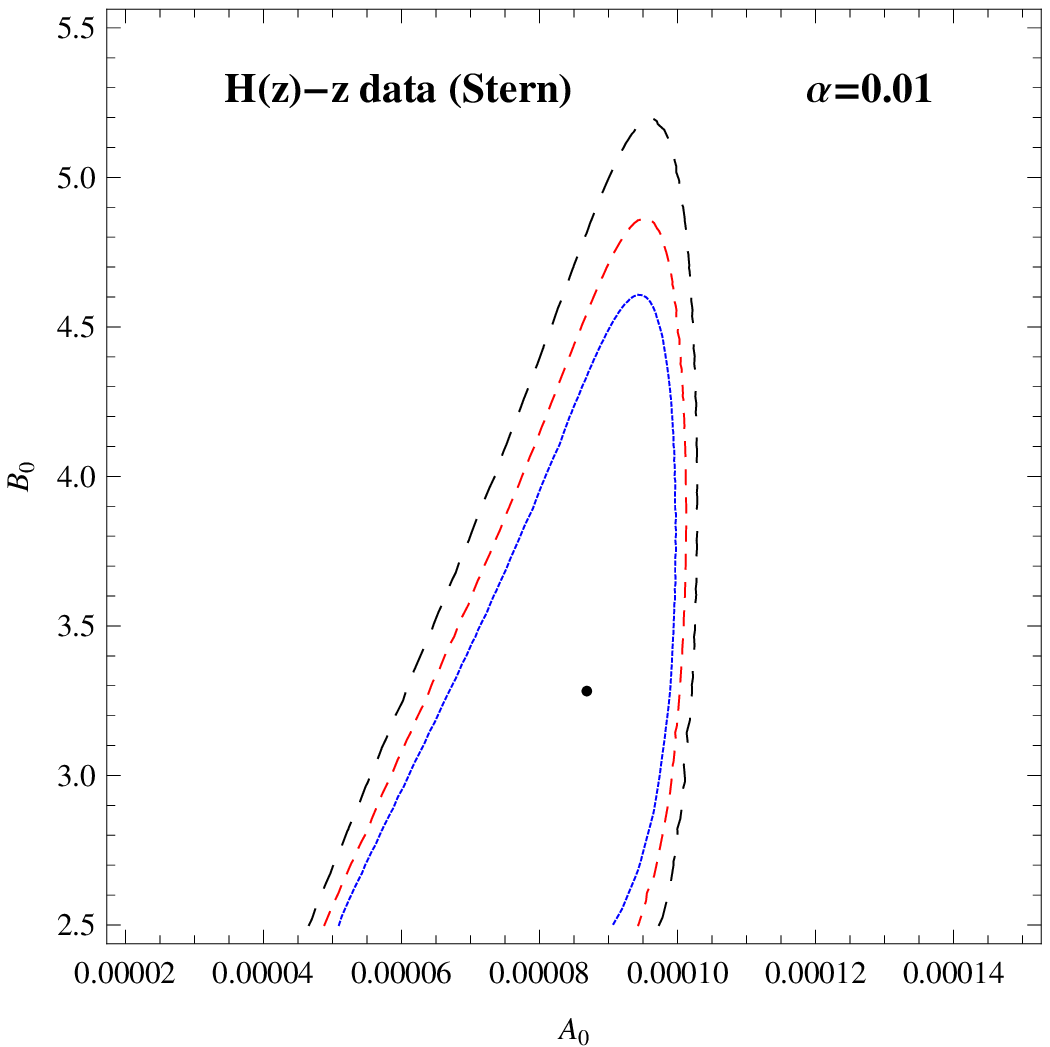}
\vspace{.2in}

~~~Fig.1~~~~~~~~~~~~~~~~~~~~~~~~~~~~~~~~~~~~~~~~~~~~~~~~~~~~~~~~~~~~~~~~~~~~~Fig.2~~~

\vspace{.2in} Figs. 1 and 2 show the variation of $A_0$ with $B_0$
for $\alpha=0.0001$ and $\alpha=0.01$ respectively for different
confidence levels. The 66\% (solid, blue), 90\% (dashed, red) and
99\% (dashed, black) contours are plotted for the $H(z)$-$z$
(Stern) analysis. \vspace{0.2in}
\end{figure}

\subsection{Analysis of $H(z)$-$z$ with BAO Peak Parameter}

In this section we use the method of joint analysis, the Baryon
Acoustic Oscillation (BAO) peak parameter proposed by Eisenstein
et al \cite{Eisenstein}. Acoustic peaks occurred because
cosmological perturbations excite sound waves in initial
relativistic plasma in the early epoch of the Universe. Sloan
Digital Sky Survey (SDSS) survey is one of the first redshift
survey (46748 luminous red galaxies spectroscopic sample) by which
the BAO signal has been directly detected at a scale  $\sim$ 100
MPc. The corresponding comoving scale of the sound horizon shell
is about 150 Mpc in radius. We shall investigate the two
parameters $A_0$ and $B_0$ for our model using the BAO peak joint
analysis for low redshift (with range $0 < z < 0.35$) using
standard $\chi^2 $ distribution. The BAO peak parameter may be
defined by

\begin{eqnarray}
{\cal{A}}=\frac{\sqrt{\Omega_m}}{E(z_1)^{1/3}}\left(\frac{\int_{0}^{z_1}\frac{dz}{E(z)}}{z_1}\right)^{2/3}
\end{eqnarray}

where
\begin{eqnarray}
\Omega_m=\Omega_m^0 (1+z_1)^3 E(z_1)^{-2}
\end{eqnarray}

Here, $E(z)$ is the normalized Hubble parameter (i.e., $E(z)\equiv
E(A_0, B_0, \alpha, m, n, C_0, z)$) and $z_1 = 0.35$ is the
typical redshift of the SDSS data sample. This quantity can be
used even for more general models which do not present a large
contribution of dark energy at early times \cite{Doran}. Now the
$\chi^2$ function for the BAO measurement can be written as in the
following form

\begin{eqnarray}
\chi^2_{BAO}=\frac{({\cal{A}}-0.469)^2}{0.017^2}
\end{eqnarray}

where the value of the parameter ${\cal{A}}$ for the flat model
($k=0$) of the FRW universe is obtained by ${\cal{A}}=0.469 \pm
0.017$ using SDSS data set \cite{Eisenstein} from luminous red
galaxies survey. Now the total joint data analysis (Stern+BAO) for
the $\chi^2$ function defined by

\begin{eqnarray}
\chi^2_{Tot}=\chi^2_{H-z}+\chi^2_{BAO}
\end{eqnarray}

\begin{figure}
\epsfxsize = 3 in \epsfysize = 2.5 in \epsfbox{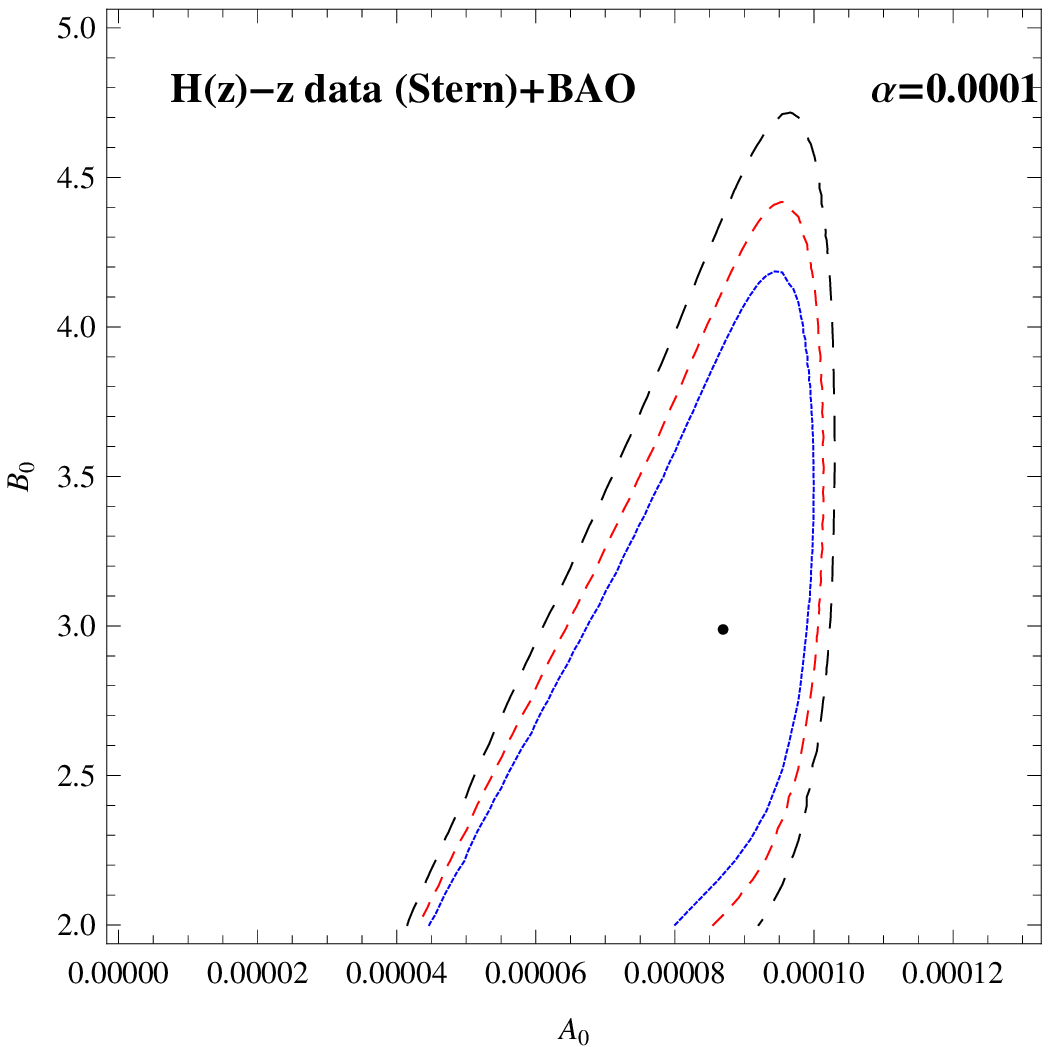}~~~~~
\epsfxsize = 3 in \epsfysize = 2.5 in \epsfbox{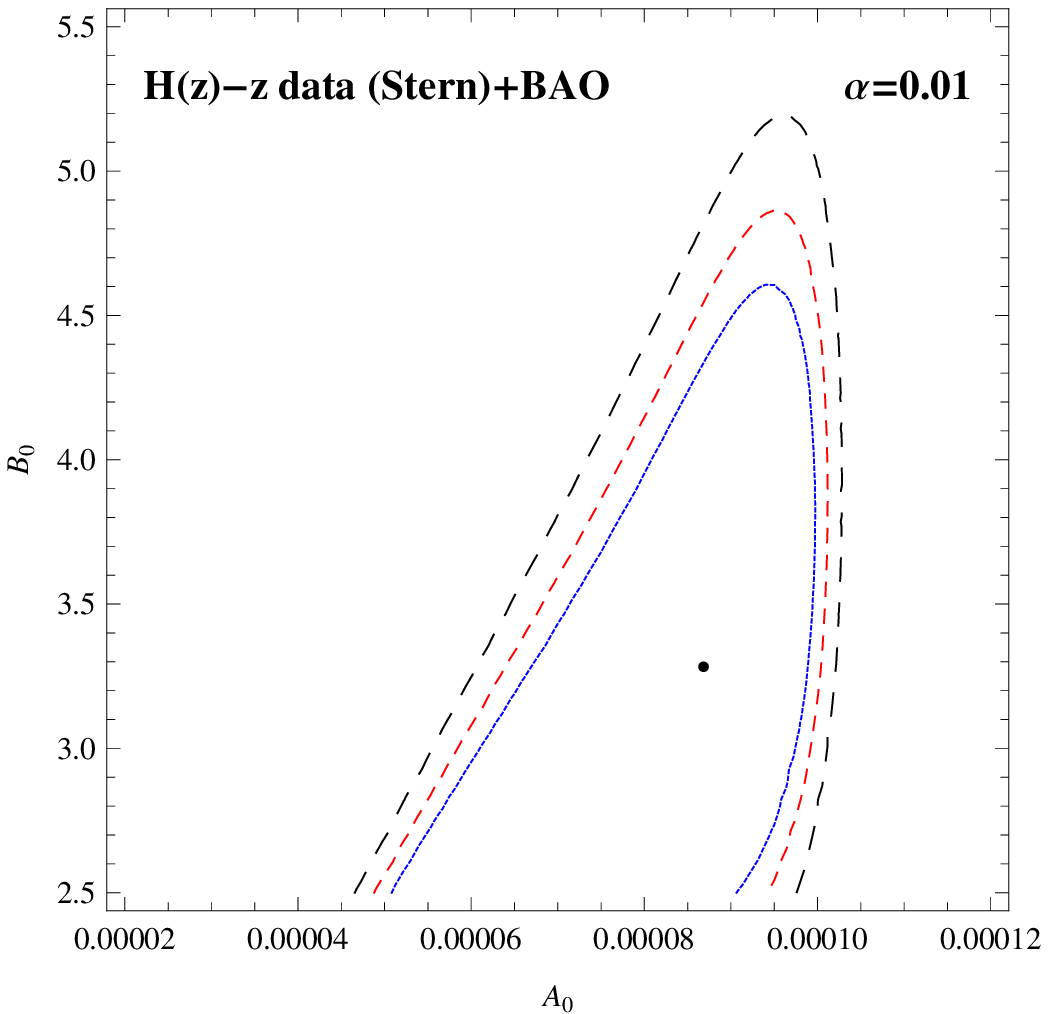}
\vspace{.2in}

~~~Fig.3~~~~~~~~~~~~~~~~~~~~~~~~~~~~~~~~~~~~~~~~~~~~~~~~~~~~~~~~~~~~~~~~~~~~~Fig.4~~~

\vspace{.2in} Figs. 3 and 4 show the variation of $A_0$ with $B_0$
for $\alpha=0.0001$ and $\alpha=0.01$ respectively for different
confidence levels. The 66\% (solid, blue), 90\% (dashed, red) and
99\% (dashed, black) contours are plotted for the $H(z)$-$z$
(Stern)+ BAO joint analysis. \vspace{0.2in}
\end{figure}

Now our best fit analysis with Stern$+$BAO observational data
support the theoretical range of the parameters. In figures 3 and
4, we plot the graphs for different confidence levels 66\% (solid,
blue), 90\% (dashed, red) and 99\% (dashed, black) contours for
$\alpha=0.0001$ and $0.01$ respectively and by fixing the other
parameters. The best fit values of $A_{0}$, $B_{0}$ and the
minimum values of $\chi^{2}$ are tabulated in Table 3.

\[
\begin{tabular}{|c|c|c|c|}
\hline
  ~~~~~~$\alpha$ ~~~~~& ~~~~~~~$A_{0}$ ~~~~~~~~& ~~~$B_{0}$~~~~~&~~~~~$\chi^{2}_{min}$~~~~~~\\
  \hline

  $0.0001$ & 0.0000868 & 2.989 & 32.165 \\
  $~~0.01$ & 0.0000867 & 3.283 & 31.924 \\
   \hline
\end{tabular}
\]
{\bf Table 3:} $H(z)$-$z$ (Stern) $+$ BAO: The best fit values of
$A_{0}$, $B_{0}$ and the minimum values of $\chi^{2}$ for $m=7$,
$n=13$, $C_0=0.1$ and for different values of $\alpha$.

\subsection{Analysis with $H(z)$-$z$, BAO Peak Parameter and CMB Shift Parameter}

Another dynamical parameter that is used in recent cosmological
tests is the CMB (Cosmic Microwave Background) shift parameter
which is the useful quantity to characterize the position of the
CMB power spectrum first peak. The CMB power spectrum first peak
is the shift parameter which is given by \cite{Elgaroy,Efstathiou}

\begin{eqnarray}
{\cal {R}}=\sqrt{\Omega_m} \int_{0}^{z_2} \frac{dz'}{H(z')/H_0}
\end{eqnarray}

where $z_2$ is the value of $z$ at the surface of last scattering.
WMAP data gives ${\cal {R}}=1.726 \pm 0.018 $ at $z=1091.3$. For
CMB measurement $\chi^2$ function can be defined as

\begin{eqnarray}
\chi^2_{CMB}=\frac{({\cal {R}}-1.726)^2}{(0.018)^2}
\end{eqnarray}

and the total joint data analysis (Stern+BAO+CMB) for the $\chi^2$
function defined by

\begin{eqnarray}
\chi^2_{Tot}=\chi^2_{H-z}+\chi^2_{BAO}+\chi^2_{CMB}
\end{eqnarray}

\begin{figure}
\epsfxsize = 3 in \epsfysize = 2.5 in \epsfbox{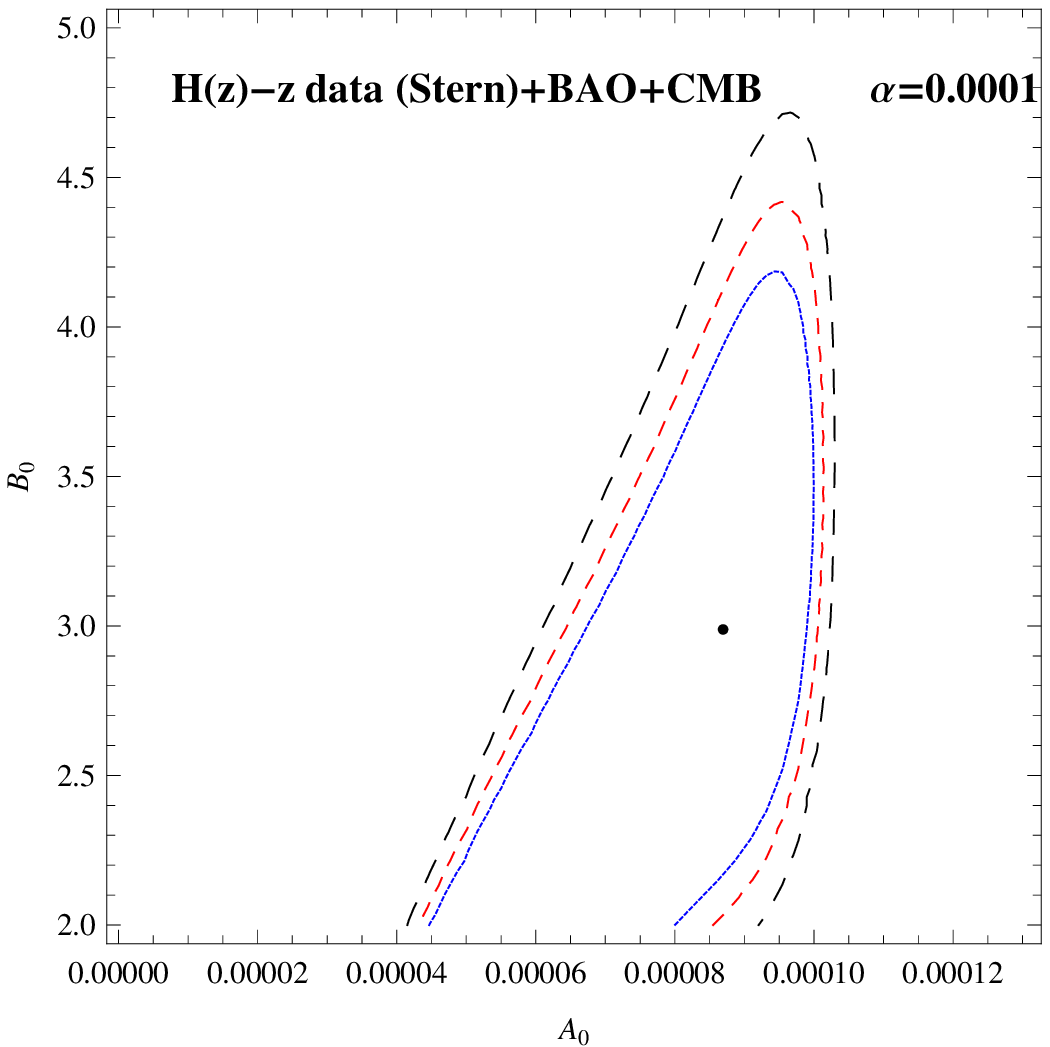}~~~~~
\epsfxsize = 3 in \epsfysize = 2.5 in \epsfbox{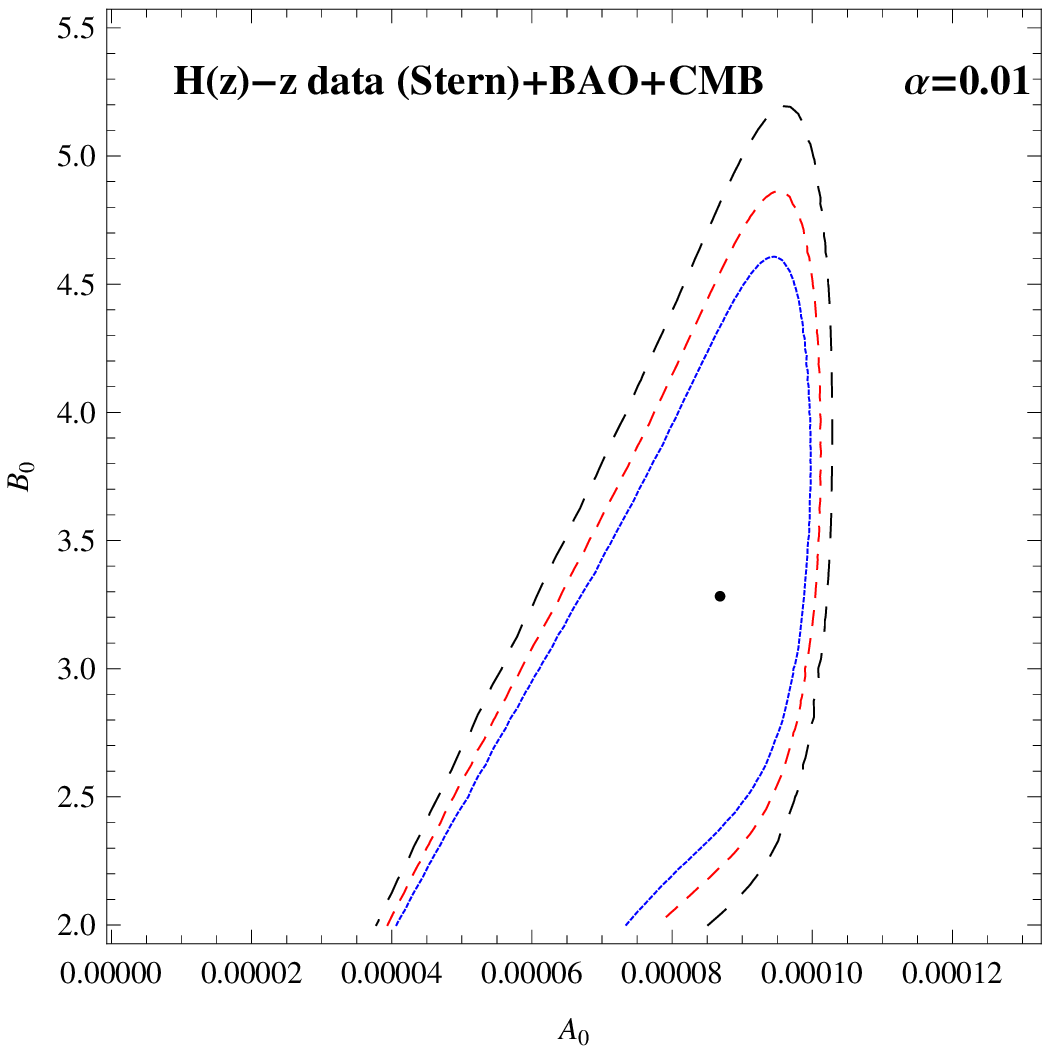}
\vspace{.2in}

~~~Fig.5~~~~~~~~~~~~~~~~~~~~~~~~~~~~~~~~~~~~~~~~~~~~~~~~~~~~~~~~~~~~~~~~~~~~~Fig.6~~~

\vspace{.2in} Figs. 5 and 6 show the variation of $A_0$ with $B_0$
for $\alpha=0.0001$ and $\alpha=0.01$ respectively for different
confidence levels. The 66\% (solid, blue), 90\% (dashed, red) and
99\% (dashed, black) contours are plotted for the $H(z)$-$z$
(Stern)+ BAO+ CMB joint analysis. \vspace{0.2in}
\end{figure}

Now our best fit analysis with Stern $+$ BAO $+$ CMB observational
data support the theoretical range of the parameters. In figures 5
and 6, we plot the graphs for different confidence levels 66\%
(solid, blue), 90\% (dashed, red) and 99\% (dashed, black)
contours for $\alpha=0.0001$ and $0.01$ respectively and by fixing
the other parameters. The best fit values of $A_{0}$, $B_{0}$ and
the minimum values of $\chi^{2}$ are tabulated in Table 4.

\[
\begin{tabular}{|c|c|c|c|}
\hline
  ~~~~~~$\alpha$ ~~~~~& ~~~~~~~$A_{0}$ ~~~~~~~~& ~~~$B_{0}$~~~~~&~~~~~$\chi^{2}_{min}$~~~~~~\\
  \hline

  $0.0001$ & 0.0000869 & 2.987 & 32.165 \\
  $~~0.01$ & 0.0000867 & 3.283 & 31.926 \\
   \hline
\end{tabular}
\]
{\bf Table 4:} $H(z)$-$z$ (Stern) $+$ BAO $+$ CMB: The best fit
values of $A_{0}$, $B_{0}$ and the minimum values of $\chi^{2}$
for $m=7$, $n=13$, $C_0=0.1$ and for different values of $\alpha$.

\begin{figure}
\epsfxsize = 3 in \epsfysize = 2.5 in \epsfbox{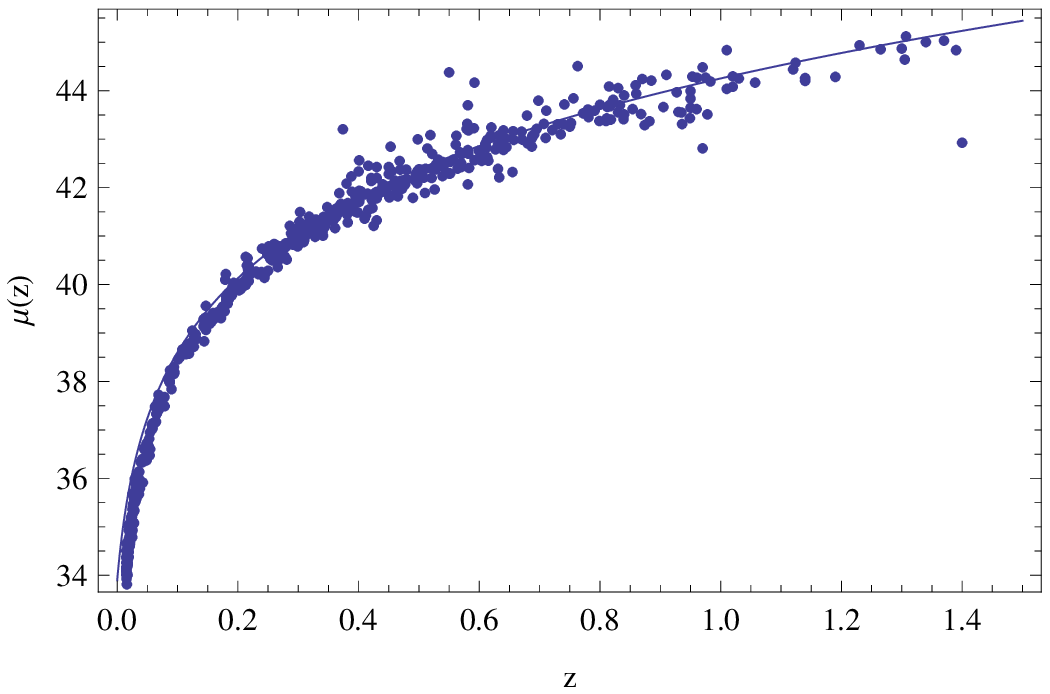}
\vspace{.2in}

~~~Fig.7~~~

\vspace{.2in} Fig.7 shows the variation of distance modulus
$\mu(z)$ vs redshift $z$ for our model (solid line) and the Union2
sample (dotted points). \vspace{0.2in}
\end{figure}

\section{Redshift-Magnitude Observations from Supernovae Type I$a$}

Recent observation of high redshift survey Supernovae Type Ia
indicates that the universe undergoing an accelerating phase and
gives an evidence of existence of dark energy. Since 1995, two
teams of High-z Supernova Search and the Supernova Cosmology
Project have discovered several type Ia supernovas at the high
redshifts \cite{Perlmutter1,Perlmutter2,Riess}. The observations
directly measure the distance modulus of a Supernovae and its
redshift $z$ \cite{Riess1}. Recent observational data, including
SNe Ia which consists of the gold sample which has a 157 supernova
\cite{Riess2} and the another set is a combined data set of a 192
supernova \cite{Riess1}. Here we consider 557 data points and
belongs to the Union2 sample \cite{Kowalaski} which are considered in the next.\\

The luminosity distance $d_L (z)$ and distance modulus $\mu(z)$
for Supernovas are calculated by
\begin{equation}
d_L(z)= (1+z) H_0 \int_{0}^{z} \frac{dz'}{H(z)}\\
\end{equation}
and
\begin{equation}
\mu(z)=5 \log_{10} \left[\frac{d_L (z)/H_0}{1 MPc}\right]+25
\end{equation}

The best fit of distance modulus as a function $\mu(z)$ of
redshift $z$ for our theoretical model and the Supernova Type Ia
Union2 sample are drawn in figure 7 for our best fit values of
$A_{0}$, $B_{0}$ with the other chosen parameters. From the
curves, we see that the theoretical NVMCG model is in
agreement with the union2 sample data.\\

\section{Discussion}

We proposed here the FRW universe filled with dark matter (perfect
fluid with negligible pressure) along with new variable Modified
Chaplygin gas (NVMCG) which is one of the candidates of dark
energy. We present the Hubble parameter $H$  in terms of the
observable parameters $\Omega_{m}^{0}$, $H_{0}$ with the redshift
$z$ and the other parameters like $A_{0}$, $B_{0}$, $C_{0}$,
$\alpha$, $m$ and $n$. We have chosen the observed values of
$\Omega_{m}^{0}=0.28$ and $H_{0}$ = 72 Kms$^{-1}$ Mpc$^{-1}$. From
Stern data set (12 points), we have obtained the bounds of the
arbitrary parameters $A_{0}$ and $B_{0}$ (Table 2) by minimizing
the $\chi^{2}$ test and by fixing the other parameters
$m=7,~n=13,~\alpha=0.1,~0.0001$ and $C_{0}=0.1$. In this way, we
may found the bounds of any two parameters by fixing the remaining
parameters. Next due to joint analysis of BAO and CMB
observations, we have also obtained the best fit values and the
bounds of the parameters ($A_{0},B_{0}$) (Table 3 and 4) by fixing
some other parameters $m=7,~n=13,~C_{0}=0.1$ and
$\alpha=0.01,~0.0001$. The best-fit values and bounds of the
parameters are obtained by 66\%, 90\% and 99\% confidence levels
are shown in figures 1-6 for Stern, Stern+BAO and Stern+BAO+CMB
analysis. The distance modulus $\mu(z)$ against redshift $z$ has
been drawn in figure 7 for our theoretical model of the NVMCG for
the best fit values of the parameters and the observed SNe Ia
Union2 data sample. Here we show that our predicted theoretical
NVMCG model permitted the observational data sets. The
observations do in fact severely constrain the nature of allowed
composition of matter-energy by constraining the range of the
values of the
parameters for a physically viable NVMCG model.\\\\

{\bf Acknowledgement:}\\

The authors are thankful to IUCAA, Pune, India for warm
hospitality where part of the work was carried out. Also UD is
thankful to CSIR, Govt. of India for providing research project
grant (No. 03(1206)/12/EMR-II).\\


\begin{thebibliography}{99}
\bibitem{Perlmutter1}Perlmutter, S. J. et al, 1998, Nature 391, 51.
\bibitem{Perlmutter2}Perlmutter, S.J. et al, 1999, Astrophys. J. 517, 565.
\bibitem{Riess}Riess, A. G. et al., 1998, Astron. J. 116, 1009.
\bibitem{Tonry}Tonry, J. L. et al., 2003, ApJ, 594, 1.
\bibitem{Melchiorri}Melchiorri A. et al.,2000,Astrophysics J.Lett.,536,L63.
\bibitem{Lange}Lange A.E. et al.,2001,Phys Rev. D, 63,042001.
\bibitem{Jaffe}Jaffe A.H. et al.,2001,Phys Rev. Lett.,86,3475.
\bibitem{Netterfield}Netterfield C.B. et al.,2002,Astrophysics,J.,571,604.
\bibitem{Halverson}Halverson N.W. et al.,2002,Astrophysics. J.,568,38.
\bibitem{Briddle}Briddle S. et al.,2003,Science,299,1532.
\bibitem{Bennet}Bennet C. et al.,2003,[astro-ph/0302207].
\bibitem{Hinshaw}Hinshaw G. et al.,2003,[astro-ph/0302217].
\bibitem{Kogut} Kogut A. et al.,2003,[astro-ph/0302213].
\bibitem{Spergel}Spergel D.N. et al.,2003,[astro-ph/0302209].
\bibitem{Kamenshchik} A. Kamenshchik et al, {\it Phys. Lett. B} {\bf 511} 265 (2001).
\bibitem{Gorini}Gorini, V., Kamenshchik, A. and Moschella, U., 2003,
Phys. Rev. D 67, 063509.
\bibitem{Benaoum} H. Benaoum, hep-th/0205140.
\bibitem{Debnath}Debnath, U., Banerjee, A. and Chakraborty, S., 2004,
Class. Quantum Grav. 21, 5609.
\bibitem{Chakraborty} W. Chakraborty and U. Debnath, Gravitation and Cosmology, {\bf 16} 223
(2010).
\bibitem{Choudhury} Choudhury, T. R. and Padmanabhan, T., 2007, Astron. Astrophys. 429, 807.
\bibitem{Lu} Lu, J. et al, 2008, Phys. Lett. B 662, 87.
\bibitem{Jun} Dao-Jun, L. and Xin-Zhou, L., 2005, Chin. Phys. Lett., 22, 1600.
\bibitem{Stern} Stern, D. et al, 2010, JCAP 1002, 008.
\bibitem{Wu} Wu, P. and Yu, H., 2007, Phys. Lett. B 644, 16.
\bibitem{Paul1} Thakur, P., Ghose, S. and Paul, B. C., 2009, Mon. Not. R. Astron. Soc. 397, 1935.
\bibitem{Paul2} Paul, B. C., Ghose, S. and Thakur, P., arXiv:1101.1360v1
  [astro-ph.CO].
\bibitem{Paul3} Paul, B. C., Thakur, P. and Ghose, S., arXiv:1004.4256v1
 [astro-ph.CO].
\bibitem{Paul4} Ghose, S., Thakur, P. and Paul, B. C., arXiv:1105.3303v1
   [astro-ph.CO].
\bibitem{Eisenstein} Eisenstein, D. J. et al, 2005, Astrophys. J. 633, 560.
\bibitem{Doran} M. Doran, S. Stern and E. Thommes, JCAP 0704, 015 (2007).
\bibitem{Elgaroy} O. Elgaroy and T. Multamaki, Astron. Astrophys. 471, 65E (2007).
\bibitem{Efstathiou} G. Efstathiou and J. R. Bond, Mon. Not. R. Ast. Soc. 304, 75 (1999).
\bibitem{Riess1}A. G. Riess et al., Astrophys. J. 659, 98 (2007); Astrophys. J. 659, 98 (2007).
\bibitem{Riess2}A. G. Riess et al., Astrophys. J. 607, 665 (2004).
\bibitem{Kowalaski} Kowalaski et al, 2008, Astrophys. J. 686, 749.

\end{thebibliography}
\end{document}